\def\arxivversion{1}
\ifdefined\arxivversion
  \documentclass[journal,onecolumn,11pt]{IEEEtran}
\else
  \documentclass[journal]{IEEEtran}
\fi
\usepackage[T1]{fontenc}
\usepackage[utf8]{inputenc}
\usepackage{microtype}
\usepackage{amsmath,amssymb}
\usepackage{booktabs}
\usepackage{graphicx}
\usepackage{float}
\usepackage{placeins}
\usepackage{multicol}
\newenvironment{widetable}{\begin{table*}[!t]}{\end{table*}}
\usepackage{cite}
\usepackage[hidelinks]{hyperref}

\newlength{\resultsfigurewidth}
\ifdefined\arxivversion
\else
\fi
\newenvironment{resultsfigure}[1][!htb]{%
  \ifdefined\arxivversion\begin{figure}[H]\else\begin{figure}[#1]\fi
}{\end{figure}}
\title{State-of-Charge Targets for Storage Participation in Two-Settlement Markets:
Reliability, Price, and Compensation Impacts}
\author{Yichen~Li\textsuperscript{1}, Alexandre~Moreira\textsuperscript{2},
Ruiwei~Jiang\textsuperscript{1}, Miguel~Heleno\textsuperscript{2},
Julie~Mulvaney-Kemp\textsuperscript{2}\\[0.5em]
{\small\textsuperscript{1}Department of Industrial and Operations Engineering,
University of Michigan, Ann Arbor, MI, USA}\\
{\small\textsuperscript{2}Lawrence Berkeley National Laboratory, Berkeley, CA, USA}}
\date{}
\begin{document}
\raggedbottom
\maketitle

\begin{abstract}
As energy storage assumes a larger role in electricity systems, market
operation must coordinate its use across time to support reliability during
periods of system stress. Minimum state-of-charge (SOC) targets provide a
means of preserving energy for these periods, but their design shapes
reliability, operating efficiency, and storage compensation. This paper
evaluates alternative SOC target designs in a California-based day-ahead
and real-time electricity market model. The comparison examines targets
based on day-ahead schedules or rolling centralized dispatch, enforced
at the individual-resource or system level, across seasons, stress
conditions, and participant information settings. The results show that
retaining more energy does not necessarily reduce load shedding: targets
that restrict discharge can limit storage support during shortages.
Under critical summer conditions, individual targets derived from rolling
centralized dispatch reduce weekly load shedding by approximately
83--84\% relative to operation without targets. Lower RT clearing costs
and average prices can coexist with higher storage compensation,
reflecting the role of day-ahead positions and real-time deviations in
determining settlement outcomes. These
findings highlight the importance of coordinating energy retention with
timely discharge and network deliverability when designing SOC requirements
for storage participation in electricity markets.
\end{abstract}


\section{Introduction}\label{sec:introduction}

As energy storage takes on a larger role in balancing electricity supply
and demand, market design must coordinate its use across time to support
system reliability. Storage can shift energy into periods of scarcity only
if earlier charging and discharging decisions leave sufficient energy
available. In sequential day-ahead (DA) and real-time (RT) markets,
changing forecasts and dispatch conditions complicate this coordination~\cite{L24}.
Differences in information and incentives between storage participants and
the system operator also affect how energy is reserved for future needs~\cite{L08}.
These considerations motivate market mechanisms that explicitly manage
stored-energy availability during critical periods.

The California Independent System Operator (CAISO) has already used minimum
state-of-charge (SOC) requirements to address this operational need. As part
of its summer 2021 readiness measures, CAISO introduced a temporary minimum
SOC constraint to reduce the risk that storage would enter critical hours
with insufficient energy to meet its DA discharge schedules. The requirement
applied to storage providing resource adequacy capacity under tight supply
conditions identified by the residual unit commitment process. CAISO reports
that the constraint was used during the September 2022 heatwave to preserve
energy for critical hours~\cite{L25}. This experience provides a direct
operational precedent for using SOC requirements to support reliability.

CAISO's policy development also recognizes the economic consequences of
retaining energy. Its 2022 Energy Storage Enhancements final proposal
included exceptional-dispatch functionality to charge storage to a specified
SOC and hold it for a defined period, together with compensation for foregone
market opportunities~\cite{L21}. Energy retention therefore connects a
system reliability objective with changes in dispatch flexibility and
participant revenues. Evaluating SOC targets requires examining these
operational and economic effects together.

The December 2025 Storage Design and Modeling discussion paper extended
this policy discussion to system-level coordination. CAISO put forward for
stakeholder consideration a mechanism that would track the system-wide
SOC trajectory established in the DA market while allowing individual
resources to deviate~\cite{L18}. This concept motivates a distinction between
preserving energy at each storage resource and preserving an aggregate
amount across the fleet. Aggregate requirements allow energy to be
redistributed among resources, while transmission constraints determine
whether the retained energy can reach the locations that need it.

These policy developments motivate a broader question: how should SOC
requirements coordinate energy retention and timely discharge to improve
storage's contribution to the system? Answering this question requires
establishing when targets reduce load shedding and when they restrict
storage support during shortages. It also requires understanding how the
information used to construct a target and the choice between individual
and system-wide enforcement affect these outcomes. The economic assessment
must then connect reliability performance with clearing costs, nodal prices,
and storage compensation.

This paper examines targets derived from DA schedules and from an auxiliary
stochastic centralized dispatch (CD) model. CD jointly schedules generation
and storage over a rolling horizon using updated operating states and
scenarios of future demand and renewable availability. The operator-side
calculation supplies SOC references for target construction; a separate
participant-side calculation supplies price forecasts used to update RT
storage bids under every strategy. CD thus connects anticipated future
conditions with current energy-management decisions. Section~\ref{sec:auxiliary-cd}
presents the detailed formulation and its two uses.

We compare individual and system-wide versions of both target sources
against a no-target benchmark in a California-based, transmission-constrained
DA--RT market model. Targets are enforced at every RT interval. Experiments
across seasons, net-load stress conditions, and participant information
settings evaluate reliability and economic outcomes together. By relating
load shedding to SOC trajectories, target constraints, and feasible
discharge, the analysis explains when energy retention supports reliability
and when it limits the timely use of storage. Comparing these mechanisms
with changes in clearing costs, nodal prices, and participant compensation
identifies the operational and economic trade-offs that should inform SOC
target design.

\section{Related Work}\label{sec:related}

\subsection{Storage Energy Management and Intertemporal Coordination}

Storage dispatch links present market decisions to future resource
availability. Research on sequential markets therefore examines how
longer-horizon information can guide shorter-horizon operation. Zhao et
al.~\cite{L01} connect a forward market and a rolling spot market through
schedules and opportunity costs, relating efficient dispatch to incentives
under perfect forecasts. Goyal and Bhattacharya~\cite{L03} consider
coordinated DA/RT markets with battery storage and demand response,
including DA-derived load-following ramp constraints. These studies provide
approaches to carrying intertemporal operating requirements across market
stages.

The usefulness of an earlier schedule also depends on how forecasts evolve.
Ghadimi and Powell~\cite{L04} study storage with rolling forecasts using a
parameterized cost function approximation to inform look-ahead decisions.
Singhal et al.~\cite{L24} review approaches ranging from participant-managed
SOC to explicit operator optimization, including DA-based terminal
constraints, terminal stored-energy values, and extended RT horizons.
These alternatives also differ in where responsibility for managing future
energy availability is assigned.
Together, these approaches connect current dispatch to consequences beyond
the immediate clearing window. They motivate comparing a fixed DA reference
with a reference updated as operating conditions and information change,
while recognizing that the resulting energy-management decisions depend
on how the reference enters RT operation.

\subsection{SOC Targets and Reliability}

Explicit SOC requirements translate anticipated future needs into energy
that must remain available. Tapia and Dvorkin~\cite{L16} introduce
chance-constrained terminal SOC requirements for energy arbitrage,
evaluating the probability of reaching an admissible energy band and the
timing of energy preservation. Storage behavior before scarcity also
depends on market conditions and participant expectations. Visvesvaran and
Mays~\cite{L08} analyze the effects of price limits, capacity penalties,
and participant beliefs on storage dispatch. Their comparison of
centralized and market-based operation highlights the role of shared
information in aligning dispatch outcomes.

These perspectives distinguish reaching a specified energy level from
providing reliable service throughout an operating period. A terminal
requirement concerns energy at a designated endpoint; an every-period
minimum SOC target also constrains when stored energy can be released.
Its system value consequently depends on both preserving energy for
future shortages and allowing discharge when demand needs support.
CAISO's SOC-retention measures and discussion of aggregate DA SOC tracking
provide policy motivation for this comparison~\cite{L21,L18}.
The relevant reliability question is how target design changes load
shedding across the operating trajectory, alongside the amount of energy
retained.

\subsection{Opportunity Costs, Market Incentives, and Storage Compensation}

The economic value of retained energy connects SOC management to prices
and bidding. Hua et al.~\cite{L09} show that settling only the binding
interval of a multi-interval RT optimization can create incentive problems
when later optimizations treat historical losses as sunk costs. Their
analysis distinguishes fixed-horizon theoretical results from rolling
implementations. Guo and Tong~\cite{L11} examine multi-period pricing under uncertainty and
intertemporal opportunity costs. For storage, Chen et al.~\cite{L10}
propose temporal locational marginal pricing that incorporates the
marginal value of an individual resource's SOC and addresses lost
opportunity costs under their modeling assumptions. Qi and Xu~\cite{L07}
link locational storage bid bounds to SOC, uncertainty, and risk
preferences through a chance-constrained multi-period formulation.

Compensation rules further shape the incentives associated with these
decisions. The CAISO Market Surveillance Committee discusses SOC
insufficiency, multi-interval losses, and storage bid-cost recovery~\cite{L23}.
FERC's January 2025 order accepts revisions intended to reduce unwarranted
storage RT bid-cost-recovery payments~\cite{L22}. Chen et al.~\cite{L15}
examine interactions between ramping procurement, pricing, and bid-cost
recovery. This literature motivates evaluating SOC requirements through
both system outcomes and participant receipts: changes in clearing costs
and prices affect the value of charging, discharging, and retained energy.
It also motivates specifying the information and bid-update procedures
used alongside each target strategy.

\subsection{Uncertainty, Network Deliverability, and Study Positioning}

The timing and location of uncertainty shape the flexibility needed by
the system. Ela and O'Malley~\cite{L12} compare multi-period scheduling
and ramp-capability constraints in terms of reliability, efficiency, and
dispatch incentives. Chen~\cite{L02} quantifies net-load ramping and
capacity uncertainty across lead times to inform system and subregional
reserve requirements. Yurdakul et al.~\cite{L06} address the network
deliverability of DA flexible-ramping awards through stochastic
optimization. Spyrou et al.~\cite{L05} coordinate flexibility options
with energy schedules to manage imbalance risk. These studies motivate
considering temporal flexibility and spatial deliverability together.

SOC targets add an energy-retention dimension to this coordination:
adequate ramping power does not establish sufficient energy for a later
interval, and adequate aggregate energy does not establish delivery to a
constrained location. Building on these perspectives, this paper compares
DA- and rolling CD-based targets at individual and system levels, enforced
in every RT interval. A common transmission-constrained DA--RT framework
connects target construction, participant information, and feasible
redispatch. The comparison evaluates load shedding, clearing costs, nodal
prices, and storage compensation across seasonal and stress conditions,
linking differences in outcomes to the preservation and release of stored
energy.

\section{Market Framework and Minimum SOC Target Design}\label{sec:framework}

% Formulation basis: literature/model_references/DetailedFormulations_v4.tex.
% Diagram reference: 2025_12_23_StorageParticipation_ori_v4.pdf, pp. 3--5.
% Common formulation: hourly participant-side CD updates RT bids in all strategies.
% Target-source and enforcement flags are independent of the bid-update procedure.
% Only minimum SOC targets enforced at every RT interval are considered.
The framework follows the sequence of decisions in a DA--RT electricity
market. The DA market schedules generation and storage using hourly
forecasts. The RT market then redispatches resources as conditions are
realized, while retaining the DA commitment of slow conventional units.
At each hourly boundary, storage participants solve an auxiliary
centralized dispatch to forecast prices and update their RT bids under
every strategy. For CD-based targets, the ISO solves a separate auxiliary
dispatch to construct SOC references. Bid updating is therefore independent
of whether an SOC target is imposed and of its source. Minimum SOC targets
connect the selected reference trajectory to actual RT clearing.

\begin{figure*}[!t]
\centering
\includegraphics[width=\textwidth,height=0.83\textheight,keepaspectratio]{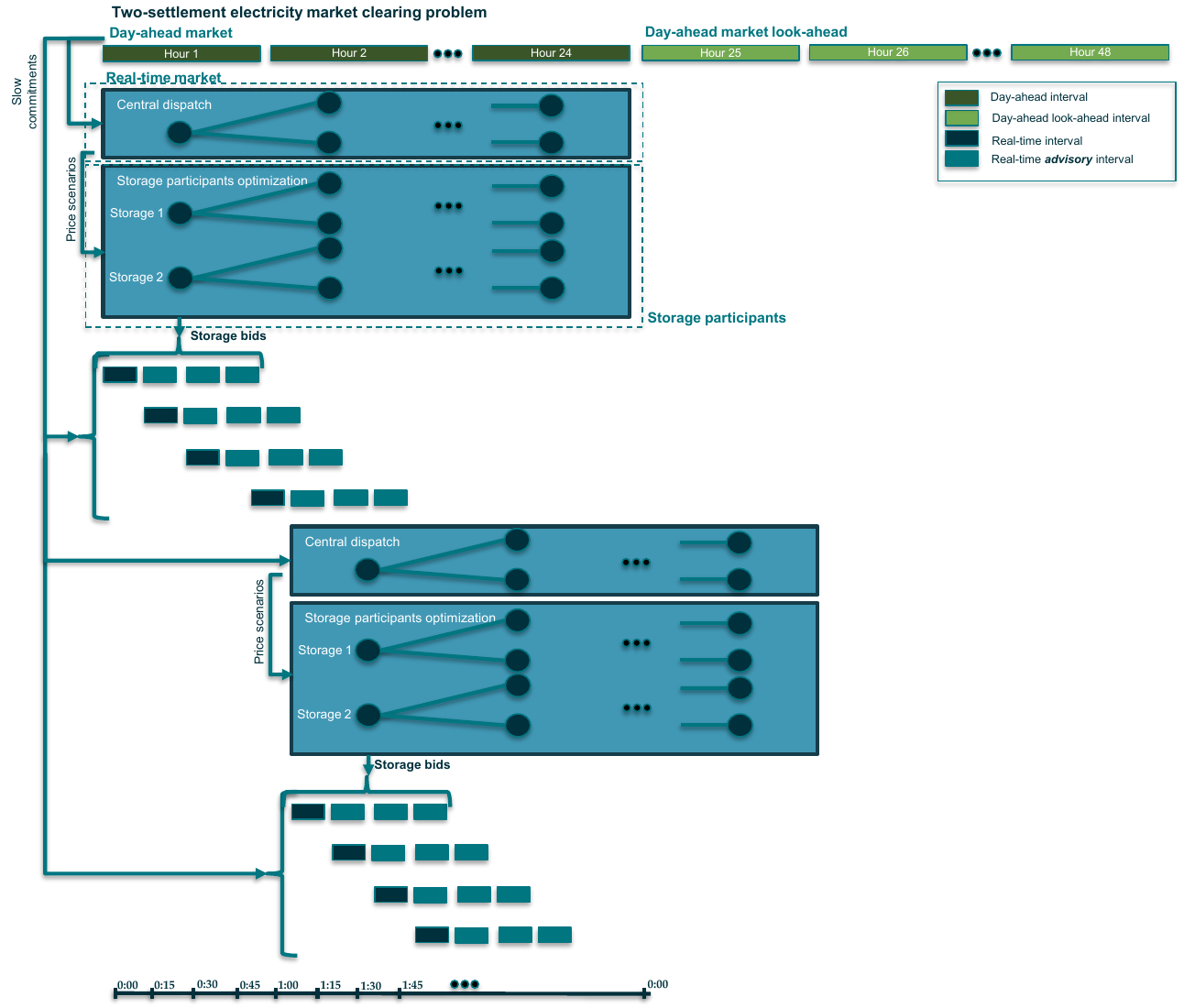}
\caption{DA--RT market framework. The upper timeline distinguishes the scheduled DA day from the DA look-ahead; the RT blocks distinguish binding and advisory intervals. The central-dispatch and storage-optimization blocks provide the participant price-forecast and bidding path, refreshed hourly under all five strategies.}\label{fig:market-framework}
\end{figure*}

Figure~\ref{fig:market-framework} presents the DA--RT framework
and its time hierarchy. Figures~\ref{fig:da-target-information}
and~\ref{fig:cd-target-information} show the DA and intraday ISO CD
information paths for target construction. The diagrams also
show the participant price-forecast and bidding path, which is refreshed
hourly for all five strategy definitions. The following
subsections describe the DA market, the RT market, the auxiliary model,
the target designs, and storage settlement and compensation in that order.
Let $\mathcal N$, $\mathcal I$,
$\mathcal K$, and $\mathcal L$ denote buses, generators, storage resources,
and transmission lines. The sets $\mathcal I_n$ and $\mathcal K_n$ contain
resources at bus $n$. Hourly periods are indexed by $h$ or $q$, and RT
intervals by $t$. The respective durations are $\Delta h=1$ hour and
$\Delta t=1/M$ hours, with $M=4$. Storage energy is measured in MWh and
power in MW. Day and year indices are suppressed.

\subsection{Day-Ahead Market}\label{sec:da-market}\label{sec:market-framework}

DA clearing establishes the schedules and energy references used in
subsequent RT operation. It uses a 48-hour hourly horizon $\mathcal H$,
covering the scheduled day and the following day's look-ahead. Inputs
include demand and renewable forecasts, resource characteristics,
inherited operating states, and storage bid coefficients constructed as
follows.

Before DA clearing, a preliminary system scheduling problem represents
storage throughput degradation cost $C_k^{\mathrm{deg}}$. After fixing its
binary decisions, the storage energy-balance dual provides the marginal
energy value $\vartheta^{\mathrm{DA}}_{k,h}$. The submitted DA bids are
\begin{equation}
\begin{aligned}
C^{D,\mathrm{DA}}_{k,h}
&=C_k^{\mathrm{deg}}+\vartheta^{\mathrm{DA}}_{k,h}/\eta_k^D,\\
C^{C,\mathrm{DA}}_{k,h}
&=C_k^{\mathrm{deg}}-\eta_k^C\vartheta^{\mathrm{DA}}_{k,h}.
\end{aligned}
\label{eq:da-bids}
\end{equation}
Here $\eta_k^C$ and $\eta_k^D$ are charging and discharging efficiencies.
Charging bids are signed objective coefficients; a negative coefficient
represents willingness to pay for charging energy.

Given these storage bids, the DA market clears energy and reserves.
For an hourly dispatch vector $x_h$, let $C_{G,h}(x_h)$ collect conventional generation, commitment, startup,
shutdown, and reserve costs. Define storage bid cost and total imbalance as
\begin{align}
C_{K,h}^{\mathrm{DA}}&=\sum_{k\in\mathcal K}
\bigl(C^{D,\mathrm{DA}}_{k,h}p^{\mathrm{dis}}_{k,h}
\notag\\
&\qquad+C^{C,\mathrm{DA}}_{k,h}p^{\mathrm{ch}}_{k,h}\bigr),
\label{eq:storage-bid-cost}\\
\mathcal I_h^{\mathrm{DA}}&=\sum_{n\in\mathcal N}
(\Delta D^-_{n,h}+\Delta D^+_{n,h})
\notag\\
&\qquad+r_h^{\mathrm{up,short}}+r_h^{\mathrm{dw,short}}.
\label{eq:imbalance-quantity}
\end{align}
With imbalance penalty $C_I$, the DA clearing problem is
\begin{equation}
\min_{x^{\mathrm{DA}}\in\mathcal X^{\mathrm{DA}}}
\sum_{h\in\mathcal H}\Delta h
\left[C_{G,h}(x_h^{\mathrm{DA}})
+C_{K,h}^{\mathrm{DA}}+C_I\mathcal I_h^{\mathrm{DA}}\right].
\label{eq:da-clearing}
\end{equation}
The feasible set $\mathcal X^{\mathrm{DA}}$ contains the hourly network,
generator, and storage constraints. Superscripts on physical dispatch
variables are omitted below when the market stage is clear.

At each bus, the hourly nodal balance is
\begin{equation}
\begin{aligned}
&\sum_{i\in\mathcal I_n}p_{i,h}
+\sum_{k\in\mathcal K_n}(p^{\mathrm{dis}}_{k,h}-p^{\mathrm{ch}}_{k,h})\\
&\quad-\sum_{\ell\in\mathcal L}A_{n\ell}f_{\ell,h}
+\Delta D^-_{n,h}-\Delta D^+_{n,h}=D_{n,h}.
\end{aligned}
\label{eq:nodal-balance}
\end{equation}
Here $A_{n\ell}$ is $1$ for a line leaving bus $n$, $-1$ for a line
entering it, and zero otherwise. Deficit and surplus variables are
nonnegative, with $\Delta D^-_{n,h}\leq D_{n,h}$. A lossless DC network
representation imposes
\begin{equation}
\begin{aligned}
f_{\ell,h}&=\frac{\theta_{o(\ell),h}-\theta_{d(\ell),h}}{x_\ell},\\
-\overline F_\ell&\leq f_{\ell,h}\leq\overline F_\ell,
\end{aligned}
\label{eq:network}
\end{equation}
with one bus angle fixed. The symbols $x_\ell$, $o(\ell)$, and $d(\ell)$
denote reactance and the two line endpoints. Upward and downward reserve
requirements are
\begin{align}
\sum_{i\in\mathcal I^{\mathrm{res}}}r^{\mathrm{up}}_{i,h}
+r_h^{\mathrm{up,short}}
&\geq R_h^{\mathrm{up,min}}\sum_nD_{n,h},\\
\sum_{i\in\mathcal I^{\mathrm{res}}}r^{\mathrm{dw}}_{i,h}
+r_h^{\mathrm{dw,short}}
&\geq R_h^{\mathrm{dw,min}}\sum_nD_{n,h},
\label{eq:reserve-requirements}
\end{align}
where $\mathcal I^{\mathrm{res}}$ contains reserve providers. Generator
output and reserves satisfy capacity, response, ramping, and minimum-time
constraints, and renewable output is bounded by forecast availability.

Hourly storage operation satisfies
\begin{align}
v^{\mathrm{DA}}_{k,h}&=v^{\mathrm{DA}}_{k,h-1}
+\Delta h\left(\eta_k^Cp^{\mathrm{ch}}_{k,h}
-p^{\mathrm{dis}}_{k,h}/\eta_k^D\right),
\label{eq:da-energy}\\
\underline V_k&\leq v^{\mathrm{DA}}_{k,h}\leq\overline V_k,
\label{eq:da-energy-limits}\\
0&\leq p^{\mathrm{ch}}_{k,h}\leq\overline P_k^{\mathrm{ch}},\quad
0\leq p^{\mathrm{dis}}_{k,h}\leq\overline P_k^{\mathrm{dis}},
\label{eq:da-storage-power}
\end{align}
The hourly system model uses independent charging and discharging bounds.
In linked daily simulations, initial storage and generator states are
inherited from the preceding implemented RT day. The corresponding
standalone cyclic formulation instead equates initial and terminal energy.

Clearing \eqref{eq:da-clearing} produces three inputs for the subsequent
market stages. Slow-unit commitment decisions carry into RT clearing;
DA schedules and nodal prices establish the DA component of settlement;
and hourly SOC trajectories provide the references for DA-target
construction. The additional minimum SOC targets studied here are imposed
in RT clearing, while the DA problem retains the physical storage bounds
specified above.

\subsection{Real-Time Market}\label{sec:rt-market}

The RT market operates on a rolling one-hour horizon comprising four
consecutive fifteen-minute intervals,
$\mathcal T_\tau=\{\tau,\tau+1,\tau+2,\tau+3\}$.
At each clearing step, the model uses the latest implemented operating
states and updated demand and renewable availability inputs to optimize
dispatch over these four intervals. Only the first interval is binding
and implemented; the remaining three intervals are advisory. The horizon
then advances by fifteen minutes, and the optimization is repeated using
the updated system state. This window extends one hour from each clearing
time, including across hourly boundaries.

Hourly bid updates and CD-based SOC targets convey information from the
auxiliary model's longer look-ahead horizon to the one-hour RT clearing
problem. The RT model applies the counterparts of
\eqref{eq:nodal-balance}--\eqref{eq:reserve-requirements} at the finer time
resolution, together with the interval-scaled generator operating limits.

For storage, define the net rate at which energy enters the inventory as
\begin{align}
\phi_{k,t}&=\eta_k^Cp^{\mathrm{ch}}_{k,t}
-p^{\mathrm{dis}}_{k,t}/\eta_k^D,
\label{eq:energy-rate}\\
v_{k,t}&=v^{\mathrm{pre}}_{k,t|\tau}+\Delta t\,\phi_{k,t},
\label{eq:energy-balance}\\
\underline V_k&\leq v_{k,t}\leq\overline V_k,
\label{eq:energy-limits}\\
0&\leq p^{\mathrm{ch}}_{k,t}\leq\overline P_k^{\mathrm{ch}}z_{k,t},
\notag\\
0&\leq p^{\mathrm{dis}}_{k,t}\leq
\overline P_k^{\mathrm{dis}}(1-z_{k,t}),\quad z_{k,t}\in\{0,1\}.
\label{eq:storage-power}
\end{align}
The binary charging mode prevents simultaneous charging and discharging
in RT. The predecessor energy is
\begin{equation}
v^{\mathrm{pre}}_{k,t|\tau}=
\begin{cases}
V^{\mathrm{DA}}_{k,0},&t=\tau=1,\\
v^{\mathrm{impl}}_{k,\tau-1},&t=\tau>1,\\
v_{k,t-1},&t>\tau,
\end{cases}
\label{eq:predecessor-state}
\end{equation}
where $V^{\mathrm{DA}}_{k,0}=v^{\mathrm{DA}}_{k,0}$ is the day's initial
energy. Thus, the binding interval starts from the implemented state,
while an advisory interval starts from the preceding decision in its own
window. Energy transitions are scaled by $\Delta t$; power bounds remain
in MW.

The principal DA--RT commitment link applies to slow conventional units.
For hour $h(t)$ containing interval $t$,
\begin{equation}
\begin{aligned}
u^{\mathrm{com}}_{i,t}&=U^{\mathrm{com,DA}}_{i,h(t)},\\
u^{\mathrm{SU}}_{i,t}&=U^{\mathrm{SU,DA}}_{i,h(t)},\\
u^{\mathrm{SD}}_{i,t}&=U^{\mathrm{SD,DA}}_{i,h(t)},
\quad i\in\mathcal I^{\mathrm{slow}}.
\end{aligned}
\label{eq:commitment-link}
\end{equation}
Fast units can adjust commitment within their operating limits. Ramping
at each window's start is anchored to implemented generation. These
physical restrictions define $\mathcal X_\tau$.

Given submitted storage bids, RT clearing solves
\begin{equation}
\min_{x\in\mathcal X_\tau\cap\mathcal Q_\tau}
\sum_{t\in\mathcal T_\tau}\Delta t
\left[C_{G,t}(x_t)+C_{K,t}+C_I\mathcal I_t\right],
\label{eq:rt-clearing}
\end{equation}
where $C_{K,t}$ and $\mathcal I_t$ are the interval counterparts of
\eqref{eq:storage-bid-cost}--\eqref{eq:imbalance-quantity}. RT upward and
downward reserve quantities include the total reserve components represented
by the formulation. The set $\mathcal Q_\tau$ contains the selected
minimum SOC constraints, defined in Section~\ref{sec:target-construction}.
No target omits these additional constraints. All strategies use the
hourly RT storage-bid updates described next, including No target and the
DA-target strategies. For these strategies, the participant-side CD informs
bidding without supplying the enforced target trajectory. Actual dispatch
and state changes are determined by this
RT clearing problem under every strategy.

\subsection{Centralized Dispatch Auxiliary Model}\label{sec:auxiliary-cd}\label{sec:information-bidding}

The auxiliary CD provides two forms of forward-looking information:
SOC references for the ISO's target construction and nodal price forecasts
for participants' RT bidding. Both calculations use the same hourly
scenario-based optimization, which jointly schedules generation and
storage under uncertain demand and renewable availability. At the start
of hour $h$, it uses the latest implemented state and current-hour realized
inputs, followed by scenario-dependent inputs over a look-ahead set
$\mathcal H_h$ of up to 24 hours, limited by available data.

As shown in Fig.~\ref{fig:cd-target-information}, the ISO-side calculation
uses scenario set $\Omega^I$, while the separate participant-side
calculation uses $\Omega^S$. These sets can differ.
Underlying weights $w_\omega$ are normalized within the selected set:
\begin{equation}
\pi_\omega^a=\frac{w_\omega}{\sum_{\nu\in\Omega^a}w_\nu},
\quad \omega\in\Omega^a,\quad a\in\{I,S\}.
\label{eq:scenario-probabilities}
\end{equation}
For either information set, the auxiliary dispatch has the compact form
\begin{equation}
\begin{aligned}
\min_{\{x_\omega\}\in\mathcal X_h^a}\quad
&\sum_{\omega\in\Omega^a}\pi_\omega^a
\sum_{q\in\mathcal H_h}\Delta h\Bigg[
C_{G,q}(x_{q,\omega})+C_I\mathcal I_{q,\omega}\\
&\qquad+\sum_{k\in\mathcal K}C_k^{\mathrm{deg}}
\left(p^{\mathrm{ch}}_{k,q,\omega}+p^{\mathrm{dis}}_{k,q,\omega}\right)
\Bigg].
\end{aligned}
\label{eq:iso-cd}
\end{equation}
The set $\mathcal X_h^a$ includes hourly network and resource constraints,
observed initial states, and first-hour nonanticipativity. In every
scenario, slow-unit commitment, startup, and shutdown also remain fixed
to the DA schedule through the hourly counterpart of
\eqref{eq:commitment-link}; fast units can adjust within their limits. Writing the
current-hour decision vector as $x_{h,\omega|h}$, first-hour
nonanticipativity requires
\begin{equation}
x_{h,\omega|h}=x_{h,\omega'|h},
\qquad \omega,\omega'\in\Omega^a.
\label{eq:cd-nonanticipativity}
\end{equation}
Later hourly decisions may adapt to each scenario. This is the stochastic
aspect of CD: future uncertainty enters a probability-weighted objective,
with a common current-hour decision and scenario-dependent recourse.
The auxiliary objective values storage throughput at degradation cost,
rather than using participant bid coefficients. Its schedules are
references for the subsequent calculations; they do not directly implement
RT dispatch or replace the DA slow-unit commitment link.

\noindent\textit{SOC references for target construction:}
For $a=I$, the output is a scenario-dependent energy trajectory
$v^{I,*}_{k,q,\omega|h}$, used to construct the CD targets in
Section~\ref{sec:target-construction}.

\noindent\textit{Price forecasts and storage bidding:}
For $a=S$,
nodal-balance duals after fixing binary decisions yield price scenarios
$\lambda^S_{n,q,\omega|h}$. After applying the bidding model's price cap,
each storage participant solves a price-taking arbitrage problem:
\begin{equation}
\begin{aligned}
\max_{y_k\in\mathcal B_{k,h}}\quad
&\sum_{\omega\in\Omega^S}\pi_\omega^S
\sum_{q\in\mathcal H_h}\Delta h\,
\lambda^S_{n(k),q,\omega|h}\\
&\qquad\times\left(p^{\mathrm{dis}}_{k,q,\omega|h}
-p^{\mathrm{ch}}_{k,q,\omega|h}\right).
\end{aligned}
\label{eq:participant-problem}
\end{equation}
Here $n(k)$ is the resource's bus, and $\mathcal B_{k,h}$ contains storage
energy dynamics, observed initial energy, physical energy and power bounds,
binary charging modes, and first-hour nonanticipativity. Network effects
enter through the supplied nodal prices.

Fixing participant binary decisions provides energy-balance multipliers
$\mu_{k,q,\omega|h}$ in the unscaled minimization of negative expected
revenue. The current-hour coefficients are
\begin{equation}
\begin{aligned}
C^{D,\mathrm{CD}}_{k,h|h}
&=\max_{\omega\in\Omega^S}
\{-\mu_{k,h,\omega|h}/\eta_k^D\},\\
C^{C,\mathrm{CD}}_{k,h|h}
&=\min_{\omega\in\Omega^S}
\{\eta_k^C\mu_{k,h,\omega|h}\}.
\end{aligned}
\label{eq:cd-bids}
\end{equation}
For later hours, multipliers are divided by $\pi_\omega^S$ before conversion
to scenario-specific coefficients, and probability-weighted bids are used
in advisory RT intervals. Current-hour bids remain constant across its
four intervals. This participant-side bidding procedure applies to every
strategy. The ISO-side reference update is used only for CD targets;
DA targets retain the DA SOC reference, and No target imposes no additional
SOC floor. The common bid-update rule separates the choice of target
source and enforcement level from the frequency of participant bidding.

\subsection{Minimum SOC Target Designs}\label{sec:target-construction}

\begin{figure*}[!t]
\centering
\includegraphics[width=\textwidth,height=0.83\textheight,keepaspectratio]{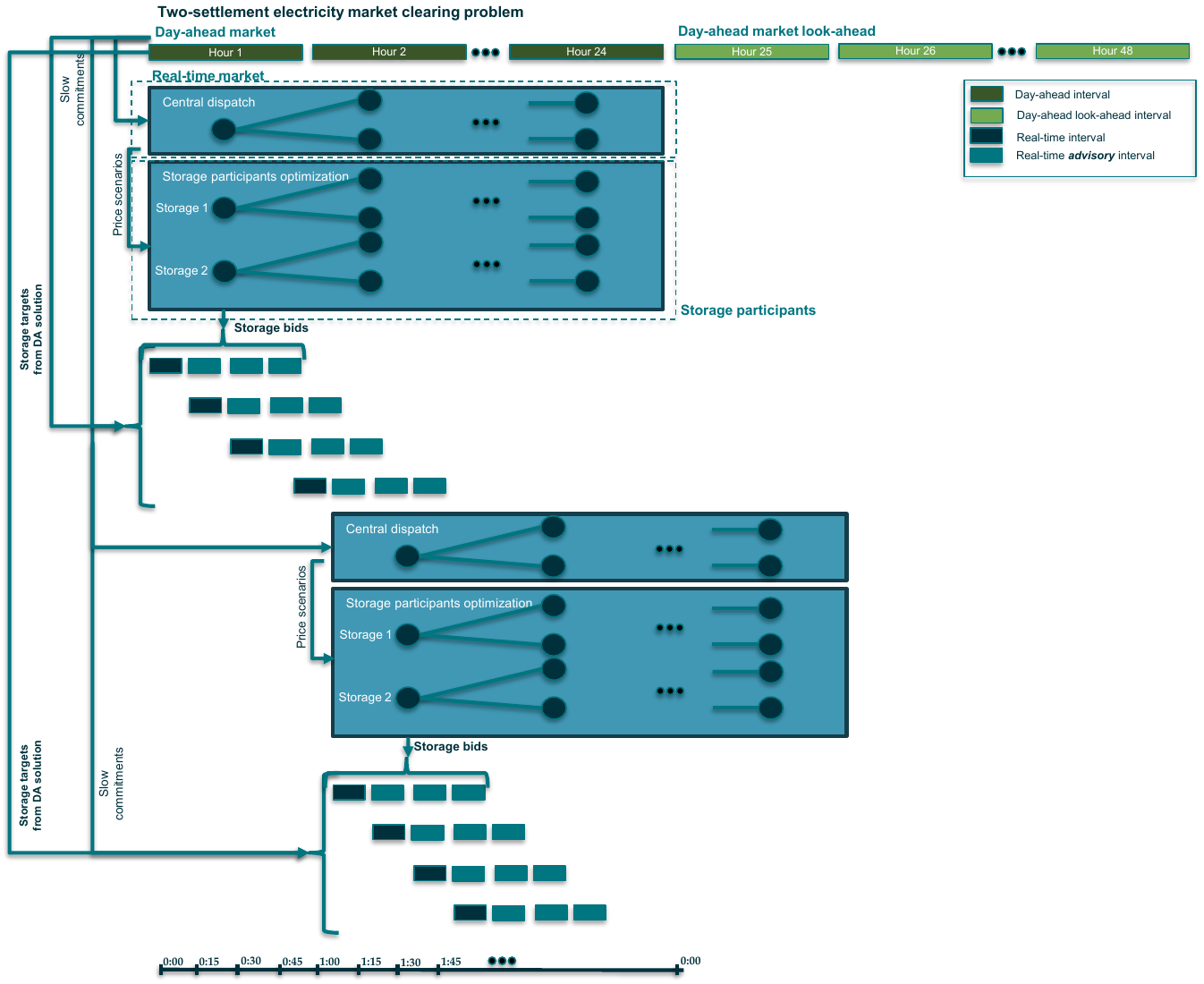}
\caption{DA-target information flow. Target information comes from the cleared DA SOC trajectory and is passed to RT clearing alongside slow-unit commitments. The individual and system floors use the same source, with enforcement defined in the text.}\label{fig:da-target-information}
\end{figure*}

\begin{figure*}[!t]
\centering
\includegraphics[width=\textwidth,height=0.83\textheight,keepaspectratio]{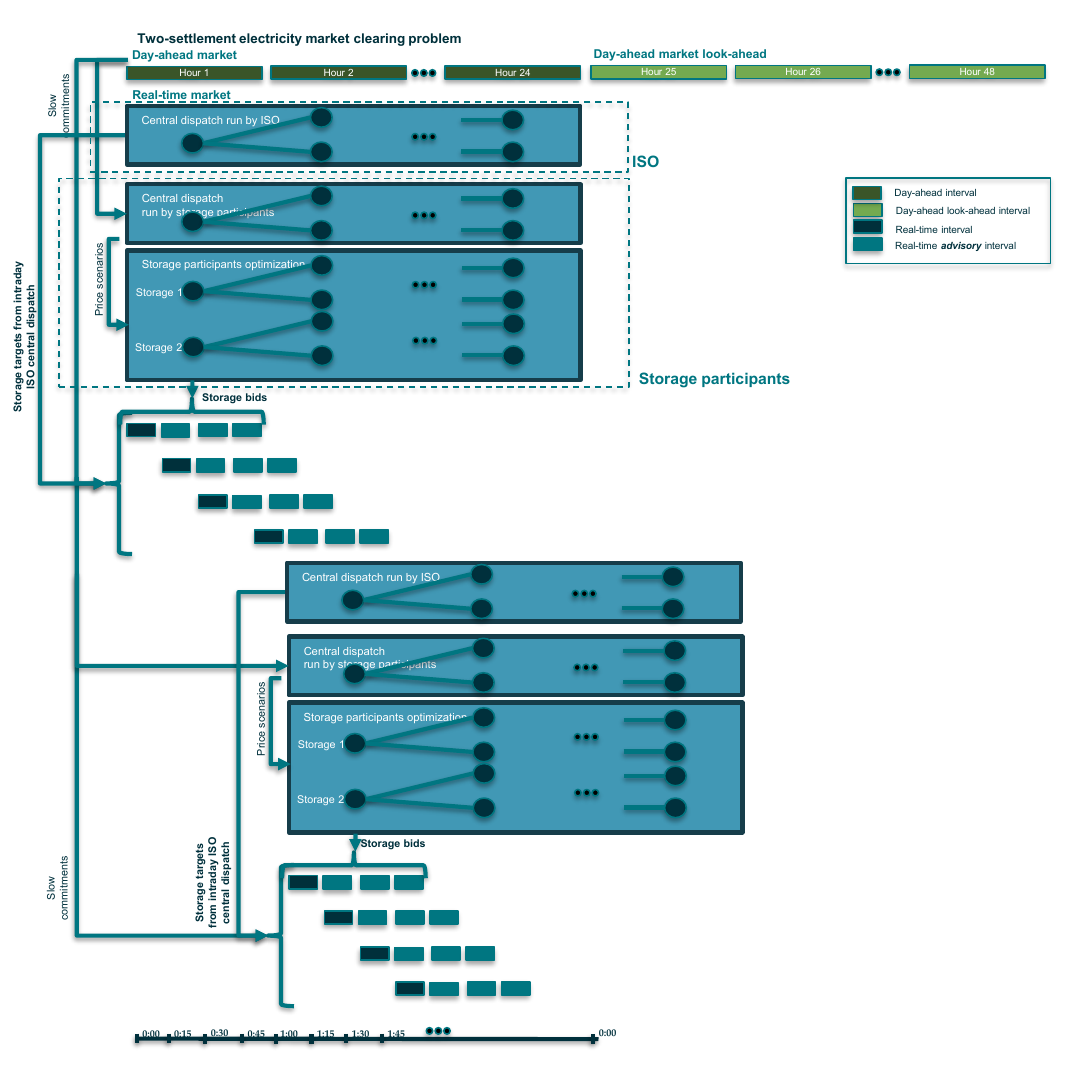}
\caption{Intraday CD-target information flow. ISO centralized dispatch supplies target information, while the participant-side centralized dispatch supplies price scenarios to the storage bidding problem. Both processes recur at hourly boundaries; RT clearing implements the binding fifteen-minute decision.}\label{fig:cd-target-information}
\end{figure*}

Figures~\ref{fig:da-target-information} and~\ref{fig:cd-target-information}
contrast the two sources of target information: the cleared DA schedule
and intraday ISO centralized dispatch. Within either source, the second
design choice is whether the resulting floor applies to each storage
resource or to the fleet. The same RT clearing
problem then enforces the selected floor at every interval. These choices
yield DA individual, DA system, CD individual, and CD system targets,
compared with the No-target benchmark.

\subsubsection{Hourly references from DA or CD}
Let $R^a_{k,q|h}$ and $R^a_{\Sigma,q|h}$ denote individual and system
references, with source $a\in\{\mathrm{DA},\mathrm{CD}\}$ and most recent
hourly update $h$. DA references use the cleared schedule and remain fixed:
\begin{equation}
\begin{aligned}
R^{\mathrm{DA}}_{k,q|h}&=v^{\mathrm{DA}}_{k,q},\\
R^{\mathrm{DA}}_{\Sigma,q|h}&=\sum_{k\in\mathcal K}v^{\mathrm{DA}}_{k,q}.
\end{aligned}
\label{eq:da-targets}
\end{equation}
For CD, the ISO's optimized trajectories from~\eqref{eq:iso-cd} give
\begin{align}
R^{\mathrm{CD}}_{k,q|h}
&=\max_{\omega\in\Omega^I}v^{I,*}_{k,q,\omega|h},
\label{eq:cd-individual-reference}\\
R^{\mathrm{CD}}_{\Sigma,q|h}
&=\max_{\omega\in\Omega^I}\sum_{k\in\mathcal K}v^{I,*}_{k,q,\omega|h}.
\label{eq:cd-system-reference}
\end{align}
Current-hour SOC is common across scenarios by nonanticipativity. For
later hours, the maxima form an upper envelope of the optimized reference
energies. Taking the maximum separately for each resource differs from
taking the maximum fleet total:
\begin{equation}
R^{\mathrm{CD}}_{\Sigma,q|h}
\leq\sum_{k\in\mathcal K}R^{\mathrm{CD}}_{k,q|h}.
\label{eq:envelope-order}
\end{equation}
The CD individual design can therefore impose a larger total energy floor
than the CD system design. These envelopes summarize the selected scenario
solutions and do not establish a reliability guarantee for other realizations.

\subsubsection{Conversion to every-period targets}
For $t=M(q-1)+m$, $m\in\{1,\ldots,M\}$, let $\alpha_t=m/M$. Linear
interpolation converts hourly references to interval energy floors:
\begin{align}
\widehat v^a_{k,t|h}
&=(1-\alpha_t)R^a_{k,q-1|h}+\alpha_tR^a_{k,q|h},
\label{eq:interpolate-individual}\\
\widehat v^a_{\Sigma,t|h}
&=(1-\alpha_t)R^a_{\Sigma,q-1|h}+\alpha_tR^a_{\Sigma,q|h}.
\label{eq:interpolate-system}
\end{align}
This rule applies to rising and falling references. For DA targets, both
endpoints stay on the DA trajectory. For a CD individual update at $h>1$,
the preceding endpoint is reset to implemented energy
$v^{\mathrm{impl}}_{k,M(h-1)}$. A CD system update replaces current and
future aggregate references while retaining the preceding aggregate
reference as the interpolation endpoint. At the day's first boundary,
initial endpoints use the day's initial energy. These reference updates
are distinct from the physical state update~\eqref{eq:energy-balance},
which always starts from implemented energy in the binding interval.

\subsubsection{Individual and system enforcement}\label{sec:target-enforcement}
For a given source $a$, the individual design adds
\begin{equation}
v^{\mathrm{pre}}_{k,t|\tau}+\Delta t\,\phi_{k,t}
\geq\widehat v^a_{k,t|h},
\quad k\in\mathcal K,\ t\in\mathcal T_\tau,
\label{eq:individual-floor}
\end{equation}
whereas the system design adds
\begin{equation}
\sum_{k\in\mathcal K}
\left(v^{\mathrm{pre}}_{k,t|\tau}+\Delta t\,\phi_{k,t}\right)
\geq\widehat v^a_{\Sigma,t|h},
\quad t\in\mathcal T_\tau.
\label{eq:system-floor}
\end{equation}
By~\eqref{eq:energy-balance}, these constraints require
$v_{k,t}\geq\widehat v^a_{k,t|h}$ or
$\sum_kv_{k,t}\geq\widehat v^a_{\Sigma,t|h}$, respectively. Both apply
to every interval in the RT window, including advisory intervals, and
allow energy above the reference. Their nonnegative multipliers,
$\gamma^a_{k,t|\tau}$ and $\gamma^a_{t|\tau}$, measure the marginal
clearing cost of tightening the floor after fixing binary decisions.
No target retains only the physical minimum $\underline V_k$.

A system requirement allows one resource to fall below its individual
reference when other resources retain sufficient energy. For DA references,
linearity of interpolation gives
$\widehat v^{\mathrm{DA}}_{\Sigma,t|h}
=\sum_k\widehat v^{\mathrm{DA}}_{k,t|h}$, so individual floors imply
the system floor at a fixed decision time, for the same initial state
and remaining constraints. This feasible-set relation does not rank
realized reliability after successive decisions and state updates.
All strategies retain nodal balances and transmission limits, so an
aggregate floor does not make energy at different buses freely transferable.

The target constraints contain no violation slack. An infeasible reference
cannot be satisfied by paying a target penalty; the imbalance variables
instead record the supply--demand consequences of operation within the
physical and target limits. After the first RT interval is implemented,
its state feeds the next window and, at the next hourly boundary, the
auxiliary calculations. This links the physical state updates to the repeated hourly and RT
sequence illustrated in Fig.~\ref{fig:market-framework}.

\subsection{Storage Settlement and Compensation}\label{sec:storage-settlement}

Storage compensation is calculated through two-settlement energy accounting.
Let $n(k)$ denote the bus of resource $k$, and define net injection as
$q_{k,t}=p^{\mathrm{dis}}_{k,t}-p^{\mathrm{ch}}_{k,t}$, with
$q^{\mathrm{DA}}_{k,h}$ defined analogously for the cleared DA schedule.
The DA schedule is settled at the DA nodal price
$\lambda^{\mathrm{DA}}_{n(k),h}$, while the deviation of implemented RT
injection from that schedule is settled at $\lambda^{\mathrm{RT}}_{n(k),t}$.
Over a reporting window, resource $k$ receives
\begin{equation}
\begin{aligned}
S_k={}&\sum_{h\in\mathcal H^{\mathrm{impl}}}\Delta h\,
\lambda^{\mathrm{DA}}_{n(k),h}q^{\mathrm{DA}}_{k,h}\\
&+\sum_{t\in\mathcal T^{\mathrm{impl}}}\Delta t\,
\lambda^{\mathrm{RT}}_{n(k),t}
\left(q_{k,t}-q^{\mathrm{DA}}_{k,h(t)}\right),
\end{aligned}
\label{eq:storage-settlement}
\end{equation}
where $h(t)$ maps each RT interval to its DA hour, $\Delta h=1$~h,
and $\Delta t=0.25$~h. The sets contain only implemented hours and
intervals, so overlapping look-ahead schedules are not settled twice.
Prices are expressed in USD/MWh and injections in MW.

Positive $S_k$ denotes a net energy-market receipt; negative $S_k$ denotes
a net payment. For example, discharging less than scheduled during a
high-price RT interval creates a negative deviation settlement even if
the resource injects positive power. Compensation therefore depends on
the DA position and RT deviations, not just on gross discharge or the
average RT price. The reported measure excludes additional SOC-target
payments, uplift, reserve revenue, degradation expense, and investment
cost. It measures energy-market settlement rather than profit.

\section{Case Study Design}\label{sec:case-design}

The case study compares the five strategies defined in
Section~\ref{sec:target-construction} through four-season experiments and
two summer sensitivity studies. The seasonal experiments cover one week
in each season under two information settings---aligned information and
limited participant information---and both critical and noncritical
realized conditions. The summer sensitivity studies examine seven levels
of demand stress and seven levels of joint demand and renewable stress
under both information settings. The results first present the summer
comparison, followed by the other seasons and the sensitivity analyses,
with a focus on load shedding, RT clearing costs, nodal prices, and storage
compensation.

\subsection{Test System and Simulation Horizons}

The California-based test system contains 75 network buses, 107 transmission
branches, and 2,525 generation units with an aggregate capacity of
88,778~MW. The storage fleet comprises 164 resources connected at 24 buses,
including batteries, pumped hydro, and long-duration storage.
Table~\ref{tab:storage-fleet} summarizes its composition.
The different storage durations and network locations allow the study to
examine both intertemporal energy management and the deliverability of
retained energy. Generation, storage, and transmission capacities are held
fixed across the experiments.

\begin{table}[!htb]
\centering
\caption{Storage fleet in the test system.}\label{tab:storage-fleet}
\footnotesize
\begin{tabular}{@{}lrrr@{}}
\toprule
Technology & Units & Power (MW) & Energy (MWh) \\
\midrule
Battery & 132 & 6,185.2 & 24,740.8 \\
Pumped hydro & 27 & 2,961.8 & 35,541.6 \\
Long-duration storage & 5 & 420.0 & 42,000.0 \\
\midrule
Total & 164 & 9,567.0 & 102,282.4 \\
\bottomrule
\end{tabular}
\end{table}

Each seasonal experiment covers seven consecutive days, as listed in
Table~\ref{tab:seasonal-windows}. DA scheduling uses hourly periods over a
48-hour horizon. RT clearing uses a rolling one-hour horizon of four
consecutive fifteen-minute intervals. It implements the first interval
and advances by fifteen minutes, yielding 96 implemented intervals per
day. The auxiliary stochastic CD is recalculated hourly with up to
24 hours of look-ahead, subject to the available input horizon; its
outputs inform bids and CD-based targets for the one-hour RT problem.
Implemented storage states and the relevant generator operating states
carry forward between successive decisions and days within each weekly
simulation. Each strategy uses the same input initialization for a given
season and stress level. The optimization models are implemented in Python
with Pyomo and solved using Gurobi.

\begin{table}[!htb]
\centering
\caption{Seasonal simulation windows. Day indices refer to the annual input series.}\label{tab:seasonal-windows}
\footnotesize
\begin{tabular}{@{}llr@{}}
\toprule
Season & Month & Seven-day window \\
\midrule
Winter & January & 15--21 \\
Spring & April & 105--111 \\
Summer & August & 223--229 \\
Fall & November & 319--325 \\
\bottomrule
\end{tabular}
\end{table}

Critical scenarios apply time-varying demand and renewable perturbations
(Table~\ref{tab:critical-parameters}). Empirical benchmarks include demand
forecast errors of approximately 2--3\% (3.73\% in peak hours) and solar/wind
errors of 2.6--2.9\%/3.6--3.9\% of installed
capacity~\cite{CAISO2023June,CAISO2023July,CAISO2023August}.
Weather-tail peak demand is 12.2\% higher in CAISO's 1-in-10 than 1-in-2
assessment~\cite{CAISO2022Assessment}; wildfire-related solar reductions
reach approximately 30\% on average and 56\% on an individual day relative
to the July baseline~\cite{EIA2020Wildfire}. Additional benchmarks include
CAISO's 10--15\% solar-based reserve requirements~\cite{CAISO2024DMM}
and an event-specific wind-model forecast error of 55\%~\cite{Terren2026Forecast}.
These benchmarks inform the selected stress magnitudes.  The resulting profiles
are applied across each seven-day simulation window.

\begin{table}[!htb]
\centering
\caption{Maximum critical-scenario adjustments relative to baseline profiles.}
\label{tab:critical-parameters}
\footnotesize
\begin{tabular}{@{}lrrr@{}}
\toprule
Season & Demand increase & Solar reduction & Wind reduction \\
\midrule
Winter & 8\% & 10\% & 7\% \\
Spring & 6\% & 10\% & 7\% \\
Summer & 11\% & 25\% & 20\% \\
Fall & 7\% & 30\% & 7\% \\
\bottomrule
\end{tabular}
\par\smallskip
\begin{minipage}{\ifdefined\arxivversion 4.8in\else\columnwidth\fi}
\footnotesize
\end{minipage}
\end{table}
\FloatBarrier

\subsection{Strategies and Information Settings}

The experiments compare five strategies: No target, DA individual, DA
system, CD individual, and CD system. The four target strategies combine
two reference sources---the cleared DA schedule and the hourly ISO-side
CD---with individual-resource or system-wide enforcement. Each applies a
minimum SOC requirement at every RT interval, using the construction in
Section~\ref{sec:target-construction}. No target retains only the physical
SOC bounds. All five strategies update RT storage bids hourly through the
participant-side CD and bidding procedure in
Section~\ref{sec:auxiliary-cd}, so the target source and enforcement rule
are the distinguishing design choices.

The scenario library distinguishes noncritical operating trajectories from
a critical trajectory representing stressed net-load conditions. The ISO
considers two noncritical trajectories and the critical trajectory in both
information settings. Under \emph{aligned information}, storage participants
consider the same three trajectories. Under \emph{limited participant
information}, participants instead consider two other noncritical
trajectories and omit the critical trajectory. The ISO uses its scenario
set to construct CD-based SOC references, while participants use theirs
to forecast prices and update RT bids. Probabilities are normalized within
each selected set as in~\eqref{eq:scenario-probabilities}.

Each information setting is evaluated under a critical realization and a
noncritical realization from its participant scenario set. For the critical
realization, both the ISO and participants include that trajectory among
their possible future scenarios under aligned information; only the ISO
includes it under limited participant information. The trajectory enters
the stochastic optimization as one possible outcome with its assigned
probability. The summer sensitivity studies extend the critical operating
conditions by varying demand stress and joint demand and renewable stress
under both information settings.

\subsection{Performance Measures and Comparison Protocol}\label{sec:performance-measures}

We evaluate reliability and economic outcomes using four measures: total
load shedding, RT clearing cost, RT nodal prices, and storage compensation.

Reliability is measured by total load shedding in MWh. For the set
$\mathcal T^{\mathrm{impl}}$ of implemented RT intervals in a comparison
window,
\begin{equation}
E^{\mathrm{shed}}=\Delta t\sum_{t\in\mathcal T^{\mathrm{impl}}}
\sum_{n\in\mathcal N}\Delta D^-_{n,t},
\label{eq:case-ens}
\end{equation}
where $\Delta t=0.25$~h. RT clearing cost sums the realized interval
costs, including generation costs, storage bid valuations, and imbalance
penalties, with interval duration incorporated.

RT nodal-price statistics report the arithmetic mean, maximum, and minimum
over all buses and implemented intervals. Storage compensation follows
the DA-plus-RT-deviation settlement in Section~\ref{sec:storage-settlement}.
We report the mean, maximum, and minimum of resource-level weekly receipts,
along with the fleet total where appropriate. The mean weights all
resources equally.

Comparisons use matching seasonal windows, stress inputs, and information
settings. Weekly statistics cover seven complete days of 96 implemented
RT intervals each; advisory intervals are excluded to avoid double counting.

\section{Results and Discussion}\label{sec:results}

The experiments evaluate how minimum SOC targets affect reliability,
RT clearing cost, nodal prices, and storage settlement. Summer spatial
stress provides the main comparison, followed by winter, spring, and
fall results and two summer sensitivity sweeps. Every reported weekly
value covers seven complete days.\footnote{For display, a case-day with
less than 1~MWh of load shedding or less than 1~MW of peak deficit is
reported as zero load shedding. This convention does not change cost totals.}
The tables distinguish aligned and
limited participant information and critical and noncritical realizations.
Prices describe bus--interval outcomes; settlement statistics describe
participant-week receipts, as defined in Section~\ref{sec:performance-measures}.

\FloatBarrier
\subsection{Summer Results}\label{sec:summer-results}\label{sec:seasonal-performance}

The summer results examine how target design affects energy retention,
release during shortages, and market outcomes over one week. Critical
and noncritical conditions are evaluated under both information settings.
The analysis first connects load shedding and clearing cost to storage
operation, then examines nodal prices, participant compensation, and
system-target feasibility.

\subsubsection{Load Shedding, Clearing Cost, and Energy Release}\label{sec:daily-mechanism}

CD individual achieves the lowest load shedding and RT clearing cost
among the complete-week strategies under critical conditions in both
information settings (Figs.~\ref{fig:summer-load-shedding}
and~\ref{fig:summer-rt-cost}). Under aligned information, it reduces
load shedding by 84.4\% and cost by 20.4\% relative to No target, reaching
3,063.22~MWh and \$109.13 million. Under limited participant information,
the corresponding reductions are 83.0\% and 18.7\%. The two comparisons
show a consistent reliability and cost advantage for CD individual
within the critical operating conditions studied.

\begin{resultsfigure}[!t]
\centering
\includegraphics[width=\resultsfigurewidth]{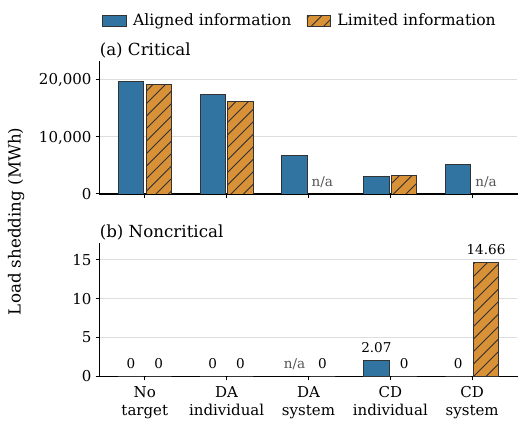}
\caption{Summer weekly load shedding under (a) critical and (b) noncritical net-load conditions. The panels use different vertical scales. Colors and hatching distinguish participant information settings; 0 denotes zero displayed load shedding and n/a an unavailable complete week.}\label{fig:summer-load-shedding}
\end{resultsfigure}

\begin{resultsfigure}[!t]
\centering
\includegraphics[width=\resultsfigurewidth]{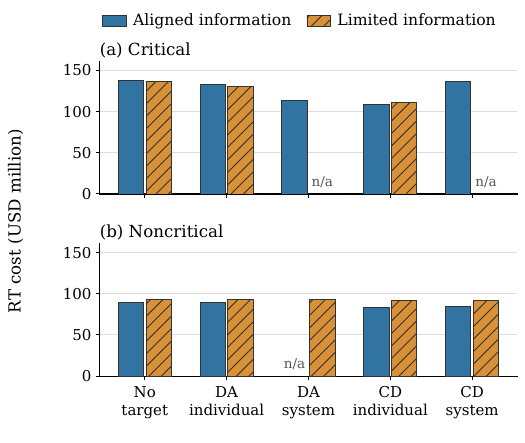}
\caption{Summer weekly RT clearing cost under (a) critical and (b) noncritical net-load conditions. Both panels use the same vertical scale. Colors and hatching distinguish participant information settings; n/a denotes an unavailable complete week.}\label{fig:summer-rt-cost}
\end{resultsfigure}

The benefits vary with target design and operating conditions. Under
aligned critical conditions, CD system reduces load shedding from
19,663.64 to 5,060.51~MWh, while cost changes only from \$137.05 million
to \$136.52 million. Reliability improvements therefore need not translate
into proportional reductions in clearing cost, which also includes
generation costs and storage bid valuations. Under aligned noncritical
conditions, CD individual lowers cost from \$89.53 million to
\$82.73 million but introduces 2.07~MWh of load shedding. Under limited
noncritical conditions, CD system introduces 14.66~MWh despite a small
cost reduction. These cases show why reliability and cost must be
evaluated together.

Greater retained energy does not necessarily translate into greater
supply during shortages. On a representative summer day under aligned
critical conditions, DA individual holds more energy than No target
during 16:00--20:00 but provides less net injection
(Table~\ref{tab:day226-mechanism}). Average SOC rises from 34.27 to
50.27~GWh, while net injection falls from 28.98 to 20.17~GWh. Load
shedding under DA individual reaches 2,293.48~MWh in this window,
almost twice the No-target value. Charging is negligible, so the lower
net injection principally reflects reduced discharge.

% Generated by tools/build_results_tables.py.
\begin{table}[!htb]
\centering
\caption{Representative summer day under aligned critical conditions. Mean SOC and net injection refer to 16:00--20:00 and are in GWh. Load shedding is in MWh; the final column covers the full day.}\label{tab:day226-mechanism}
\footnotesize
\setlength{\tabcolsep}{4pt}
\begin{tabular}{@{}lrrrr@{}}
\toprule
Strategy & Mean SOC & Injection & \shortstack{Window\\shedding} & \shortstack{Day\\shedding} \\
\midrule
No target & 34.27 & 28.98 & 1,158.52 & 2,130.59 \\
DA individual & 50.27 & 20.17 & 2,293.48 & 2,293.54 \\
CD individual & 35.47 & 26.16 & 1,004.10 & 1,092.47 \\
\bottomrule
\end{tabular}
\end{table}

The next-period target determines how much of the stored energy can be
released. For resource $k$ discharging in interval $t+1$, the floor
$\widehat v_{k,t+1}$ gives the energy-based bound
\begin{equation}
p^{\mathrm{dis}}_{k,t+1}\leq\frac{\eta_k^D}{\Delta t}
\left(v^{\mathrm{impl}}_{k,t}-\widehat v_{k,t+1}\right),
\label{eq:target-release-bound}
\end{equation}
provided the floor is at least the physical minimum. Without a target,
the corresponding bound uses $\underline V_k$. The energy margin is
\begin{equation}
\begin{aligned}
v^{\mathrm{impl}}_{k,t}-\widehat v_{k,t+1}
&=\left(v^{\mathrm{impl}}_{k,t}-\widehat v_{k,t}\right)\\
&\quad+\left(\widehat v_{k,t}-\widehat v_{k,t+1}\right).
\end{aligned}
\label{eq:target-release-margin}
\end{equation}
When SOC is at its current floor, only a decline in the next floor
permits discharge. A flat floor blocks discharge despite a positive
inventory, while a floor above current SOC requires charging. Power
and network limits can restrict release further.

At 18:00, DA individual holds 53.32~GWh against summed individual floors
of 53.18~GWh, leaving only 136.55~MWh above those floors. Of the resources
with targets above 1~MWh, 158 are within 0.01~MWh of their respective
floors. Their subsequent release consequently depends primarily on the
decline in the target trajectory. Slack at another resource cannot relax
an individual floor. Across the full day, DA individual records
2,293.54~MWh of load shedding versus 2,130.59~MWh under No target; CD individual
records 1,092.47~MWh. These results show that the timing of target
relaxation, together with the retained inventory, shapes storage
support during shortages.

\FloatBarrier
\subsubsection{RT Nodal Prices}

Lower average RT prices coexist with extreme prices under
critical conditions (Table~\ref{tab:summer-prices}). All four target
strategies lower the aligned mean relative to No target, with CD
individual producing the lowest value: \$72.84/MWh versus
\$127.36/MWh. Nevertheless, the maximum remains near \$2,000/MWh
and the minimum near zero under all five strategies. CD individual
also has the lowest mean among the complete-week limited critical
strategies, at \$68.19/MWh versus \$104.92/MWh under No target.
The reduction in average prices therefore does not eliminate extreme
prices over the reporting week.

% Generated by tools/build_results_tables.py.
\begin{table}[!htb]
\centering
\caption{Summer RT nodal prices (USD/MWh). Mean, maximum, and minimum cover all bus--interval observations; the mean is unweighted. A dash denotes an unavailable complete week.}\label{tab:summer-prices}
\footnotesize
\setlength{\tabcolsep}{4pt}
\begin{tabular}{@{}lrrr@{}}
\toprule
Strategy & Mean & Max & Min \\
\midrule
\multicolumn{4}{l}{\textit{Critical, aligned information}} \\
No target & 127.36 & 2,001.14 & 0.00 \\
DA individual & 91.33 & 2,000.00 & 0.00 \\
DA system & 92.17 & 2,000.00 & 0.00 \\
CD individual & 72.84 & 2,000.00 & 0.00 \\
CD system & 79.68 & 2,000.00 & 0.00 \\
\midrule
\multicolumn{4}{l}{\textit{Critical, limited information}} \\
No target & 104.92 & 2,001.12 & 0.00 \\
DA individual & 91.28 & 2,000.00 & 0.00 \\
DA system & --- & --- & --- \\
CD individual & 68.19 & 2,000.00 & 0.00 \\
CD system & --- & --- & --- \\
\midrule
\multicolumn{4}{l}{\textit{Noncritical, aligned information}} \\
No target & 53.15 & 98.64 & 0.00 \\
DA individual & 52.48 & 76.58 & 0.00 \\
DA system & --- & --- & --- \\
CD individual & 53.23 & 2,000.00 & 0.00 \\
CD system & 52.76 & 105.65 & 0.00 \\
\midrule
\multicolumn{4}{l}{\textit{Noncritical, limited information}} \\
No target & 54.04 & 120.10 & 0.00 \\
DA individual & 53.57 & 82.29 & 0.00 \\
DA system & 56.76 & 2,000.00 & 0.00 \\
CD individual & 52.18 & 98.64 & 0.00 \\
CD system & 59.02 & 2,000.00 & 0.00 \\
\bottomrule
\end{tabular}
\end{table}

Under noncritical conditions, similar average prices can accompany
markedly different maxima. For aligned conditions, the CD-individual
mean is \$53.23/MWh, close to
the No-target value of \$53.15/MWh, but the maximum rises from
\$98.64/MWh to \$2,000/MWh alongside the small amount of load shedding.
Under limited
noncritical conditions, CD system raises the mean to \$59.02/MWh and
the maximum to \$2,000/MWh, whereas CD individual lowers the mean to
\$52.18/MWh and limits the maximum to \$98.64/MWh. A favorable weekly
average alone does not establish favorable price extremes.

\FloatBarrier
\subsubsection{Storage Compensation}

Lower average RT prices can accompany higher storage compensation.
Under aligned critical conditions, CD individual combines the lowest
mean RT price with the highest mean participant settlement
(Table~\ref{tab:summer-settlement}). Its fleet receipts are
\$17.21 million, or \$104.93 thousand per resource on average,
compared with a fleet net payment of \$21.91 million and a mean of
$-\$133.59$ thousand under No target. Under the two-settlement rule
in Section~\ref{sec:storage-settlement}, compensation depends on the DA
position and the quantity and price of RT deviations. Average RT price
alone therefore does not determine the ordering of participant receipts.

% Generated by tools/build_results_tables.py.
\begin{table}[!htb]
\centering
\caption{Summer storage compensation (USD thousand). Mean, maximum, and minimum are across 164 participant-week energy settlements. A dash denotes an unavailable complete week.}\label{tab:summer-settlement}
\footnotesize
\setlength{\tabcolsep}{4pt}
\begin{tabular}{@{}lrrr@{}}
\toprule
Strategy & Mean & Max & Min \\
\midrule
\multicolumn{4}{l}{\textit{Critical, aligned information}} \\
No target & -133.59 & 645.61 & -1,373.73 \\
DA individual & 28.30 & 384.29 & -1.06 \\
DA system & -2.01 & 697.16 & -648.47 \\
CD individual & 104.93 & 2,183.13 & -11.62 \\
CD system & 23.19 & 590.18 & -695.96 \\
\midrule
\multicolumn{4}{l}{\textit{Critical, limited information}} \\
No target & -97.47 & 383.15 & -2,305.35 \\
DA individual & 28.15 & 397.32 & 0.04 \\
DA system & --- & --- & --- \\
CD individual & 129.68 & 2,635.28 & -4.69 \\
CD system & --- & --- & --- \\
\midrule
\multicolumn{4}{l}{\textit{Noncritical, aligned information}} \\
No target & 10.83 & 173.42 & -14.94 \\
DA individual & 17.94 & 248.88 & -2.24 \\
DA system & --- & --- & --- \\
CD individual & 15.20 & 165.53 & -15.70 \\
CD system & 6.69 & 115.17 & -78.91 \\
\midrule
\multicolumn{4}{l}{\textit{Noncritical, limited information}} \\
No target & 9.10 & 144.32 & -42.10 \\
DA individual & 18.43 & 234.64 & -2.10 \\
DA system & 17.10 & 223.46 & -23.29 \\
CD individual & 18.17 & 239.53 & -20.90 \\
CD system & -7.32 & 88.04 & -216.75 \\
\bottomrule
\end{tabular}
\end{table}

The participant extrema reveal uneven receipts across resources. Under aligned critical
conditions, CD individual ranges from $-\$11.62$ thousand to
\$2,183.13 thousand, while CD system ranges from $-\$695.96$ thousand
to \$590.18 thousand. Positive aggregate receipts do not imply that
every resource receives a positive settlement. DA individual
under limited critical conditions has a positive minimum, approximately
\$0.04 thousand, but a lower fleet total than CD individual.
The resources differ in capacity, location, and DA positions, so these
extrema describe absolute participant outcomes rather than returns per
unit of installed capacity.

In noncritical conditions, the distinction between system cost and
participant receipts remains visible. CD system lowers aligned RT cost
relative to No target, yet its fleet settlement falls from \$1.78 million
to \$1.10 million. Under limited noncritical conditions, its fleet
settlement becomes negative at $-\$1.20$ million. Target design therefore
affects the distribution of market receipts as well as operating cost
and reliability.

\FloatBarrier
\subsubsection{System-Target Feasibility}\label{sec:system-feasibility}

System-target designs have fewer complete summer trajectories than the
individual designs (Table~\ref{tab:target-coverage}). No target, DA
individual, and CD individual each have four complete summer weeks;
DA system has two and CD system has three. These counts describe the
coverage of the weekly comparisons. Unavailable weeks receive no
load-shedding or cost value; their absence alone does not identify the
cause of a failed run.

% Generated by tools/build_results_tables.py.
\begin{table}[!htb]
\centering
\caption{Complete seven-day results across seasons. The four operating conditions combine aligned or limited participant information with critical or noncritical net load. Coverage is not a solver-infeasibility classification.}\label{tab:target-coverage}
\footnotesize
\setlength{\tabcolsep}{4pt}
\begin{tabular}{@{}lrrrrr@{}}
\toprule
Strategy & Summer & Winter & Spring & Fall & Total \\
\midrule
No target & 4/4 & 4/4 & 4/4 & 4/4 & 16/16 \\
DA individual & 4/4 & 4/4 & 3/4 & 4/4 & 15/16 \\
DA system & 2/4 & 0/4 & 0/4 & 0/4 & 2/16 \\
CD individual & 4/4 & 4/4 & 4/4 & 4/4 & 16/16 \\
CD system & 3/4 & 3/4 & 4/4 & 4/4 & 14/16 \\
\bottomrule
\end{tabular}
\end{table}

System aggregation relaxes the allocation of stored energy across
resources, but it does not remove the need to reach the prescribed energy
level. The following necessary one-step condition provides a test of
aggregate target reachability:
\begin{equation}
\widehat v_{\Sigma,t+1}\leq
\sum_{k\in\mathcal K}\min\left\{\overline V_k,
v^{\mathrm{impl}}_{k,t}+\Delta t\eta_k^C\overline P_k^{\mathrm{ch}}\right\}.
\label{eq:aggregate-reachability}
\end{equation}
The right-hand side is an optimistic upper bound that assumes charging
power is available at every storage location. Network limits and supply
availability can reduce the reachable inventory. A floor exceeding this
bound is infeasible even before those restrictions are imposed. A floor
below it still requires sufficient supply and network access to achieve
the necessary charging. This condition identifies a mechanism for
infeasibility; it does not assign that mechanism to the unavailable weeks
in Table~\ref{tab:target-coverage}. Because the target is a hard constraint,
load-shedding variables do not relax it.

For a two-storage illustration, consider resources A and B, each with
10~MWh of capacity and unit charging efficiency. Their charging limits
are 16 and 4~MW, respectively, allowing at most 4 and 1~MWh of additional
energy in a 15-minute interval. Assume sufficient supply and no network
restrictions. At an hourly target update, suppose the implemented state
is $(v_{A,t},v_{B,t})=(10,0)$~MWh, satisfying the current 10~MWh system
floor. If the updated floor for the next interval is 14~MWh, the maximum
reachable total is only
$\min\{10,10+4\}+\min\{10,0+1\}=11$~MWh: A is already full, and B
cannot charge fast enough. By contrast, the state $(5,5)$~MWh has the
same total energy but can reach
$\min\{10,5+4\}+\min\{10,5+1\}=15$~MWh. It can meet the 14~MWh
floor by charging A to 9~MWh while leaving B at 5~MWh. Thus, identical
aggregate SOC can imply different feasibility for an updated target;
unused capacity must be available at resources that can charge quickly
enough.

For the same state and a DA aggregate floor equal to the sum of individual
floors, individual feasibility implies aggregate feasibility. The
aggregate constraint is therefore more flexible at a fixed decision time.
Across successive RT decisions, however, the two designs can distribute
energy differently across resources and carry different states into
later intervals. Those states determine which future targets are
reachable. System-target design must consequently account for both the
aggregate energy requirement and the evolving distribution of storage
energy and charging capability.

\FloatBarrier
\subsection{Four-Season Results}\label{sec:other-seasons}

Table~\ref{tab:seasonal-comparison} compares load shedding, RT clearing
cost, maximum nodal price, and mean storage compensation across all four
seasons under aligned critical conditions. It presents the four strategies
with complete weeks in every season. Additional winter, spring, and fall
results appear in Appendix~\ref{app:seasonal-results}.

% Generated by tools/build_results_tables.py.
\begin{table}[!htb]
\centering
\caption{Four-season comparison under aligned critical conditions. Strategies with complete weeks in all four seasons are shown. Load shedding is in MWh, RT cost in USD million, maximum nodal price in USD/MWh, and mean participant compensation in USD thousand.}\label{tab:seasonal-comparison}
\footnotesize
\setlength{\tabcolsep}{4pt}
\begin{tabular}{@{}lrrrr@{}}
\toprule
Strategy & \shortstack{Load\\shedding} & \shortstack{RT\\cost} & \shortstack{Max\\price} & \shortstack{Mean\\comp.} \\
\midrule
\multicolumn{5}{l}{\textit{Winter}} \\
No target & 1,701.27 & 61.50 & 2,000.00 & 19.69 \\
DA individual & 2,389.07 & 61.94 & 2,000.00 & 12.78 \\
CD individual & 103.79 & 57.07 & 2,000.00 & 8.23 \\
CD system & 0.00 & 57.18 & 135.71 & 8.59 \\
\midrule
\multicolumn{5}{l}{\textit{Spring}} \\
No target & 0.00 & 19.72 & 2,000.00 & 21.35 \\
DA individual & 382.16 & 21.10 & 2,000.00 & 12.56 \\
CD individual & 0.00 & 19.33 & 2,000.00 & 9.32 \\
CD system & 0.00 & 19.46 & 2,000.00 & 26.05 \\
\midrule
\multicolumn{5}{l}{\textit{Summer}} \\
No target & 19,663.64 & 137.05 & 2,001.14 & -133.59 \\
DA individual & 17,385.51 & 133.18 & 2,000.00 & 28.30 \\
CD individual & 3,063.22 & 109.13 & 2,000.00 & 104.93 \\
CD system & 5,060.51 & 136.52 & 2,000.00 & 23.19 \\
\midrule
\multicolumn{5}{l}{\textit{Fall}} \\
No target & 0.00 & 49.89 & 2,000.00 & 6.64 \\
DA individual & 0.00 & 49.47 & 2,000.00 & 7.36 \\
CD individual & 0.00 & 48.99 & 136.83 & 10.68 \\
CD system & 0.00 & 48.45 & 109.20 & 6.98 \\
\bottomrule
\end{tabular}
\end{table}

Target performance varies across seasons. CD individual has the lowest
load shedding and cost in summer, while CD system eliminates winter load
shedding and lowers the maximum nodal price. DA individual increases load
shedding in winter and introduces it in spring. Fall strategies have equal
load-shedding outcomes but differ in cost, price extremes, and storage
compensation. The income effects also vary: both CD designs improve
winter reliability while reducing mean storage receipts relative to No
target. These comparisons show that reliability, cost, and participant
income need to be assessed jointly across operating conditions.

\FloatBarrier
\subsection{Sensitivity Analysis}\label{sec:stress-sensitivity}

The sensitivity analysis compares No target, DA individual, and CD
individual across seven stress levels under both information settings.
Generation, storage, and transmission capacities remain fixed. The two
experiments assess how demand stress and joint demand--renewable stress
change weekly load shedding and RT clearing cost.

\subsubsection{Increasing Demand Stress}

Demand stress increases the configured peak-demand uplift from 0\% to
12\% in 2-percentage-point steps. Each level adjusts the within-day
load profile and its spatial distribution. Figure~\ref{fig:demand-sensitivity}
summarizes the resulting reliability and cost outcomes.

\begin{resultsfigure}[!htb]
\centering
\includegraphics[width=\resultsfigurewidth]{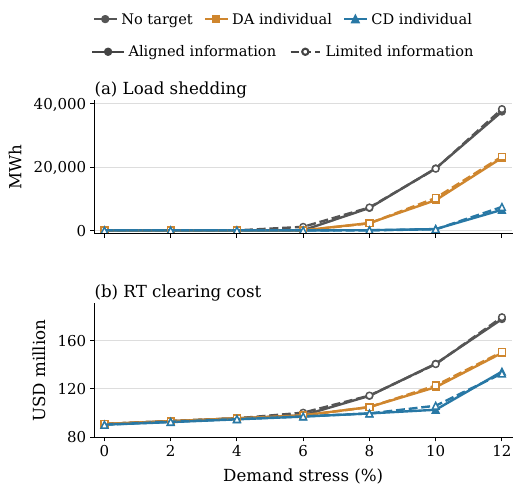}
\caption{Demand-stress sensitivity of weekly load shedding and RT clearing cost. Colors and markers identify strategies; solid/filled and dashed/open lines distinguish aligned and limited participant information.}\label{fig:demand-sensitivity}
\end{resultsfigure}

CD individual's advantage is most pronounced at stress levels of 8\%
and above. Under aligned information, it reduces load shedding by
99.2\%, 97.9\%, and 82.6\% at 8\%, 10\%, and 12\%, respectively,
with corresponding RT cost reductions of 12.7\%, 27.0\%, and 24.5\%.
The limited-information setting shows the same strong reliability gains,
with load shedding reductions of 99.6\%, 98.2\%, and 80.7\%.
At low stress, the reliability ordering varies: the aligned baseline
has 14.47~MWh under CD individual versus 1.55~MWh under No target.
At 12\%, CD individual still records 6,507.21~MWh under aligned
information and 7,408.78~MWh under limited information. Energy management
therefore delivers substantial high-stress benefits while residual
shortages grow at the upper end of the sweep.

\FloatBarrier
\subsubsection{Joint Demand and Renewable Stress}

The joint sweep varies nodal demand and solar/wind availability together
between native RT conditions and a critical reference. Stress levels are
calibrated to the reference summer day's RT--DA peak-net-load deviation,
with the resulting adjustment applied across the week. This net-load
measure differs from the peak-demand uplift in the first sweep.
Figure~\ref{fig:joint-sensitivity} presents the joint-stress results.

\begin{resultsfigure}[!htb]
\centering
\includegraphics[width=\resultsfigurewidth]{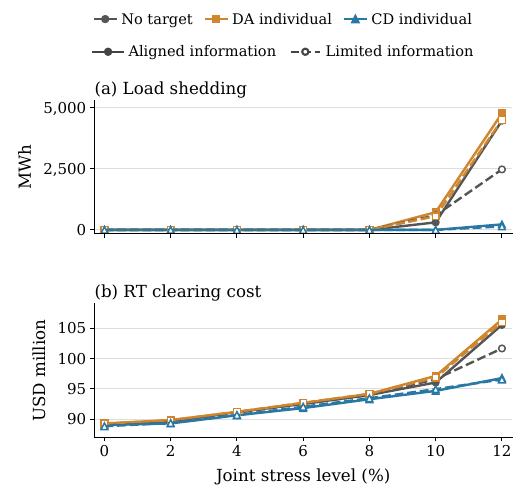}
\caption{Joint demand/renewable sensitivity of weekly load shedding and RT clearing cost. The level is calibrated to the reference-day RT--DA peak-net-load deviation. Line conventions match Fig.~\ref{fig:demand-sensitivity}.}\label{fig:joint-sensitivity}
\end{resultsfigure}

CD individual eliminates No-target load shedding at 10\% joint stress
in both information settings. At 12\%, it reduces load shedding by
94.9\% under aligned information and 94.2\% under limited information,
with RT cost reductions of 8.3\% and 5.0\%, respectively. DA individual
instead produces more load shedding than No target in both settings at
12\%, reversing its benefit in the high-demand sweep. The comparison
shows that target performance depends on the form of net-load stress:
CD individual substantially reduces severe shortages in both sweeps,
while DA-target reliability benefits vary with the demand and renewable
trajectory.

\FloatBarrier
\section{Conclusion}\label{sec:conclusion}

This paper compares minimum SOC targets derived from DA schedules and
rolling centralized dispatch in a California-based electricity market
model. The results demonstrate the value of updating energy-retention
requirements as operating conditions evolve. CD individual maintains
RT clearing costs close to or below those of No target at low stress,
while substantially reducing both load shedding and cost at high stress.
Under critical summer conditions, it reduces weekly load shedding by
83--84\% and RT clearing cost by 19--20\% across the two information
settings. The demand and joint demand--renewable sensitivity studies
reinforce this advantage under severe stress.

System targets provide flexibility in allocating stored energy across
resources, but an aggregate energy requirement does not ensure that
energy is available at the locations and times where it is needed.
Successive dispatch decisions change the distribution of SOC and the
charging capability available to meet subsequent targets. An aggregate
floor can therefore become unreachable when charging power, available
supply, or network access is insufficient. System-target design must
account for this evolving resource-level state as well as total stored
energy.

These findings support using SOC targets as adaptive operating
requirements that coordinate energy preservation with timely release.
Targets should be refreshed with updated system conditions, allow stored
energy to be discharged during shortages, and be checked against physical
and network constraints before enforcement. Raising a minimum SOC floor
alone does not improve reliability: an overly restrictive trajectory can
withhold energy precisely when it is needed. Target evaluation should
also consider nodal prices and storage compensation alongside load
shedding and clearing cost, since system benefits and participant
receipts need not move together.

\FloatBarrier
\onecolumn
\appendices
\section{Complete Seasonal Results}\label{app:seasonal-results}

Tables~\ref{tab:winter-results}--\ref{tab:fall-results} report complete-week
seasonal results. Load shedding is in MWh, RT cost in USD million, and
nodal prices in USD/MWh. Price statistics cover all bus--interval observations;
compensation statistics cover 164 participant-week settlements, in USD thousand
(Section~\ref{sec:performance-measures}). A dash denotes an unavailable complete week.

\begingroup
\setlength{\intextsep}{6pt}
\renewcommand{\arraystretch}{0.97}
% Full-width appendix tables follow the one-column document flow.
\renewenvironment{widetable}{\begin{table}[H]}{\end{table}}
% Generated by tools/build_results_tables.py.
\begin{widetable}
\centering
\caption{Winter weekly results.}\label{tab:winter-results}
\footnotesize
\setlength{\tabcolsep}{4pt}
\begin{tabular}{@{}lrrrrrrrr@{}}
\toprule
Strategy & \shortstack{Load shedding\\(MWh)} & RT cost & Price mean & Price max & Price min & Settl. mean & Settl. max & Settl. min \\
\midrule
\multicolumn{9}{l}{\textit{Aligned, critical}} \\
No target & 1,701.27 & 61.50 & 53.84 & 2,000.00 & 3.99 & 19.69 & 796.16 & -134.16 \\
DA individual & 2,389.07 & 61.94 & 53.66 & 2,000.00 & 0.00 & 12.78 & 129.24 & -32.12 \\
DA system & --- & --- & --- & --- & --- & --- & --- & --- \\
CD individual & 103.79 & 57.07 & 53.00 & 2,000.00 & 0.00 & 8.23 & 122.05 & -167.66 \\
CD system & 0.00 & 57.18 & 47.46 & 135.71 & 23.21 & 8.59 & 121.75 & -92.57 \\
\midrule
\multicolumn{9}{l}{\textit{Aligned, noncritical}} \\
No target & 0.00 & 55.26 & 47.96 & 78.59 & 27.86 & 2.67 & 59.67 & -65.51 \\
DA individual & 0.00 & 54.18 & 44.26 & 146.02 & 27.86 & 8.51 & 107.39 & -75.08 \\
DA system & --- & --- & --- & --- & --- & --- & --- & --- \\
CD individual & 0.00 & 53.95 & 44.74 & 86.67 & 24.00 & 6.94 & 125.84 & -43.06 \\
CD system & 0.00 & 53.36 & 46.14 & 99.69 & 23.77 & 3.52 & 84.65 & -135.34 \\
\midrule
\multicolumn{9}{l}{\textit{Limited, noncritical}} \\
No target & 0.00 & 56.77 & 48.60 & 101.49 & 27.86 & 3.09 & 59.67 & -53.01 \\
DA individual & 0.00 & 55.69 & 47.16 & 104.52 & 27.44 & 8.45 & 111.34 & -53.70 \\
DA system & --- & --- & --- & --- & --- & --- & --- & --- \\
CD individual & 0.00 & 56.11 & 53.82 & 2,000.00 & 22.28 & 16.17 & 233.17 & -246.28 \\
CD system & 0.00 & 56.29 & 48.05 & 165.45 & 22.23 & 4.13 & 82.53 & -93.94 \\
\midrule
\multicolumn{9}{l}{\textit{Limited, critical}} \\
No target & 1,652.06 & 61.47 & 54.09 & 2,000.00 & 4.88 & 15.06 & 1,226.25 & -407.24 \\
DA individual & 2,161.24 & 61.56 & 59.50 & 2,000.00 & 0.00 & 18.61 & 268.92 & -9.06 \\
DA system & --- & --- & --- & --- & --- & --- & --- & --- \\
CD individual & 129.07 & 57.51 & 50.30 & 2,000.00 & 21.52 & 15.36 & 172.06 & -68.72 \\
CD system & --- & --- & --- & --- & --- & --- & --- & --- \\
\bottomrule
\end{tabular}
\end{widetable}

% Generated by tools/build_results_tables.py.
\begin{widetable}
\centering
\caption{Spring weekly results.}\label{tab:spring-results}
\footnotesize
\setlength{\tabcolsep}{4pt}
\begin{tabular}{@{}lrrrrrrrr@{}}
\toprule
Strategy & \shortstack{Load shedding\\(MWh)} & RT cost & Price mean & Price max & Price min & Settl. mean & Settl. max & Settl. min \\
\midrule
\multicolumn{9}{l}{\textit{Aligned, critical}} \\
No target & 0.00 & 19.72 & 25.20 & 2,000.00 & 1.41 & 21.35 & 554.43 & -1.76 \\
DA individual & 382.16 & 21.10 & 29.15 & 2,000.00 & 0.69 & 12.56 & 221.37 & -4.94 \\
DA system & --- & --- & --- & --- & --- & --- & --- & --- \\
CD individual & 0.00 & 19.33 & 25.07 & 2,000.00 & 0.38 & 9.32 & 99.67 & -60.39 \\
CD system & 0.00 & 19.46 & 25.30 & 2,000.00 & 4.61 & 26.05 & 757.89 & -2.33 \\
\midrule
\multicolumn{9}{l}{\textit{Aligned, noncritical}} \\
No target & 0.00 & 18.67 & 24.32 & 124.09 & 0.00 & 7.02 & 96.44 & -5.23 \\
DA individual & 0.00 & 18.61 & 24.52 & 75.76 & 2.47 & 6.06 & 94.22 & -1.59 \\
DA system & --- & --- & --- & --- & --- & --- & --- & --- \\
CD individual & 0.00 & 18.21 & 24.35 & 70.43 & 0.00 & 7.76 & 98.99 & -20.50 \\
CD system & 0.00 & 17.94 & 23.86 & 91.55 & 0.00 & 4.17 & 43.88 & -32.66 \\
\midrule
\multicolumn{9}{l}{\textit{Limited, noncritical}} \\
No target & 0.00 & 19.10 & 24.41 & 81.89 & 0.00 & 6.93 & 95.80 & -4.98 \\
DA individual & 0.00 & 18.98 & 24.52 & 97.94 & 7.32 & 4.93 & 90.08 & -5.08 \\
DA system & --- & --- & --- & --- & --- & --- & --- & --- \\
CD individual & 0.00 & 18.71 & 25.03 & 85.26 & 0.00 & 7.35 & 99.51 & -20.25 \\
CD system & 0.00 & 19.22 & 23.73 & 88.45 & 0.00 & 3.28 & 85.96 & -27.70 \\
\midrule
\multicolumn{9}{l}{\textit{Limited, critical}} \\
No target & 0.00 & 19.72 & 24.91 & 2,000.00 & 0.00 & 14.33 & 270.77 & -2.08 \\
DA individual & --- & --- & --- & --- & --- & --- & --- & --- \\
DA system & --- & --- & --- & --- & --- & --- & --- & --- \\
CD individual & 0.00 & 19.85 & 26.32 & 2,000.00 & 0.38 & 12.81 & 128.36 & -4.07 \\
CD system & 0.00 & 20.35 & 24.41 & 2,000.00 & 1.75 & 21.49 & 656.93 & -5.54 \\
\bottomrule
\end{tabular}
\end{widetable}

% Generated by tools/build_results_tables.py.
\begin{widetable}
\centering
\caption{Fall weekly results.}\label{tab:fall-results}
\footnotesize
\setlength{\tabcolsep}{4pt}
\begin{tabular}{@{}lrrrrrrrr@{}}
\toprule
Strategy & \shortstack{Load shedding\\(MWh)} & RT cost & Price mean & Price max & Price min & Settl. mean & Settl. max & Settl. min \\
\midrule
\multicolumn{9}{l}{\textit{Aligned, critical}} \\
No target & 0.00 & 49.89 & 41.02 & 2,000.00 & 0.00 & 6.64 & 107.47 & -34.39 \\
DA individual & 0.00 & 49.47 & 47.54 & 2,000.00 & 0.00 & 7.36 & 83.66 & -13.96 \\
DA system & --- & --- & --- & --- & --- & --- & --- & --- \\
CD individual & 0.00 & 48.99 & 38.84 & 136.83 & 0.00 & 10.68 & 148.57 & -11.00 \\
CD system & 0.00 & 48.45 & 40.69 & 109.20 & 0.00 & 6.98 & 100.97 & -43.14 \\
\midrule
\multicolumn{9}{l}{\textit{Aligned, noncritical}} \\
No target & 0.00 & 47.40 & 40.21 & 86.62 & 0.95 & 4.71 & 69.43 & -37.77 \\
DA individual & 0.00 & 46.83 & 37.82 & 101.02 & 23.12 & 5.02 & 73.85 & -13.02 \\
DA system & --- & --- & --- & --- & --- & --- & --- & --- \\
CD individual & 0.00 & 46.91 & 40.53 & 1,977.07 & 0.00 & 7.77 & 82.77 & -17.39 \\
CD system & 0.00 & 45.86 & 39.27 & 98.27 & 12.69 & 1.86 & 68.87 & -85.19 \\
\midrule
\multicolumn{9}{l}{\textit{Limited, noncritical}} \\
No target & 0.00 & 48.98 & 41.03 & 76.31 & 0.00 & 5.05 & 99.55 & -38.66 \\
DA individual & 0.00 & 48.43 & 39.78 & 107.28 & 0.00 & 7.87 & 108.70 & -3.21 \\
DA system & --- & --- & --- & --- & --- & --- & --- & --- \\
CD individual & 0.00 & 48.40 & 42.75 & 2,000.00 & 0.00 & 8.02 & 101.04 & -8.19 \\
CD system & 0.00 & 48.36 & 39.73 & 117.11 & 11.56 & -2.87 & 31.42 & -116.13 \\
\midrule
\multicolumn{9}{l}{\textit{Limited, critical}} \\
No target & 0.00 & 49.80 & 40.32 & 96.14 & 0.00 & 5.88 & 117.43 & -22.51 \\
DA individual & 0.00 & 49.45 & 45.17 & 2,000.00 & 0.00 & 7.59 & 101.16 & -9.74 \\
DA system & --- & --- & --- & --- & --- & --- & --- & --- \\
CD individual & 0.00 & 49.50 & 39.65 & 111.61 & 0.00 & 10.11 & 119.52 & -16.57 \\
CD system & 0.00 & 49.24 & 40.50 & 119.03 & 0.00 & 0.12 & 57.55 & -148.09 \\
\bottomrule
\end{tabular}
\end{widetable}

\endgroup

\ifdefined\arxivversion

\else
\begin{multicols}{2}

\end{multicols}
\fi

\begin{thebibliography}{99}

\bibitem{L24}
Nikita Singhal, Alexandre Moreira, Julie Mulvaney Kemp, Erik Ela, Miguel Heleno,
``Navigating modeling frontiers for electric storage resources in wholesale electricity markets,''
Oct. 2024. Lawrence Berkeley National Laboratory and EPRI, technical report.

\bibitem{L08}
Siva Visvesvaran, Jacob Mays,
``Market-Based Storage Dispatch and Electricity Reliability,''
Working paper, June 2025.

\bibitem{L25}
California ISO,
``Decision on Minimum State of Charge Extension,''
Memorandum to the ISO Board of Governors, Mar. 15, 2023, pp. 1--3.
\href{https://www.caiso.com/documents/decisiononminimumstateofchargeextension-memo-mar2023.pdf}{[Online]}.

\bibitem{L21}
California ISO,
``Energy Storage Enhancements --- Final Proposal,''
Oct. 27, 2022. California ISO, final proposal, pp. 6, 16--18.
\href{https://stakeholdercenter.caiso.com/InitiativeDocuments/FinalProposal-EnergyStorageEnhancements.pdf}{[Online]}.

\bibitem{L18}
California ISO,
``Storage Design and Modeling --- Uplift and Default Energy Bid Working Group --- Updated Discussion and Issue Paper,''
Dec. 12, 2025. California ISO, discussion and issue paper, p. 27.
\href{https://stakeholdercenter.caiso.com/InitiativeDocuments/Storage-Design-and-Modeling-Updated-Discussion-and-Issue-Paper-on-Uplift-and-Default-Energy-Bid-Dec-12-2025.pdf}{[Online]}.

\bibitem{L01}
Jinye Zhao, Tongxin Zheng, Eugene Litvinov,
``A Multi-Period Market Design for Markets With Intertemporal Constraints,''
2020. IEEE Transactions on Power Systems, vol. 35, no. 4. DOI: 10.1109/TPWRS.2019.2963022.

\bibitem{L03}
Anshul Goyal, Kankar Bhattacharya,
``Design of Multi-Settlement Electricity Markets Considering Demand Response and Battery Energy Storage Systems Participation,''
2024. IEEE Transactions on Energy Markets, Policy and Regulation, vol. 2, no. 2. DOI: 10.1109/TEMPR.2024.3350510.

\bibitem{L04}
Saeed Ghadimi, Warren B. Powell,
``Stochastic search for a parametric cost function approximation: Energy storage with rolling forecasts,''
2024. European Journal of Operational Research, vol. 312, pp. 641--652.

\bibitem{L16}
Tomás Tapia, Yury Dvorkin,
``Reachability Guarantees for Energy Arbitrage,''
arXiv:2601.12081v2, Sept. 2026.

\bibitem{L09}
Bowen Hua, Dane A. Schiro, Tongxin Zheng, Ross Baldick, Eugene Litvinov,
``Pricing in Multi-Interval Real-Time Markets,''
2019. IEEE Transactions on Power Systems, vol. 34, no. 4. DOI: 10.1109/TPWRS.2019.2891541.

\bibitem{L11}
Ye Guo, Lang Tong,
``Pricing Multi-period Dispatch Under Uncertainty,''
2018. Allerton Conference on Communication, Control, and Computing.

\bibitem{L10}
Cong Chen, Lang Tong, Ye Guo,
``Pricing Energy Storage in Real-time Market,''
arXiv:2101.10151v1, Jan. 2021.

\bibitem{L07}
Ning Qi, Bolun Xu,
``Locational Energy Storage Bid Bounds for Facilitating Social Welfare Convergence,''
Manuscript, May 2025.

\bibitem{L23}
James Bushnell, Scott M. Harvey, Benjamin F. Hobbs,
``Opinion on Storage Bid Cost Recovery,''
Nov. 1, 2024. California ISO Market Surveillance Committee, final opinion.

\bibitem{L22}
Federal Energy Regulatory Commission,
``Order Accepting Tariff Revisions,''
Jan. 24, 2025. 190 FERC paragraph 61,047, Docket ER25-576-000.

\bibitem{L15}
Cong Chen, Valentina Norambuena, Lang Tong,
``Ramping Procurement and Bid-Cost Recovery in Real-Time Market,''
Working paper.

\bibitem{L12}
Erik Ela, Mark O'Malley,
``Scheduling and Pricing for Expected Ramp Capability in Real-Time Power Markets,''
IEEE Transactions on Power Systems, doi: 10.1109/TPWRS.2015.2461535.

\bibitem{L02}
Yonghong Chen,
``Addressing Uncertainties Through Improved Reserve Product Design,''
2023. IEEE Transactions on Power Systems, vol. 38, no. 4. DOI: 10.1109/TPWRS.2022.3200697.

\bibitem{L06}
Ogün Yurdakul, Erik Ela, Farhad Billimoria,
``Flexible Ramping Product Procurement in Day-Ahead Markets,''
2025. IEEE Transactions on Energy Markets, Policy and Regulation, vol. 3, no. 1. DOI: 10.1109/TEMPR.2024.3453907.

\bibitem{L05}
Elina Spyrou, Qiwei Zhang, Robin B. Hytowitz, Benjamin F. Hobbs, Siddharth Tyagi, Mengmeng Cai, Michael Blonsky,
``Flexibility Options: A Proposed Product for Managing Imbalance Risk,''
2025. IEEE Transactions on Energy Markets, Policy and Regulation, vol. 3, no. 3, doi: 10.1109/TEMPR.2025.3529689.

\bibitem{CAISO2023June}
California ISO,
``Summer Market Performance Report: June 2023,''
July 28, 2023, pp. 22--23.
\href{https://www.caiso.com/documents/summermarketperformancereportforjune2023.pdf}{[Online]}.

\bibitem{CAISO2023July}
California ISO,
``Summer Market Performance Report: July 2023,''
2023, pp. 31--33.
\href{https://www.caiso.com/documents/summer-market-performance-report-for-july-2023.pdf}{[Online]}.

\bibitem{CAISO2023August}
California ISO,
``Summer Market Performance Report: August 2023,''
Oct. 10, 2023, pp. 23--25.
\href{https://www.caiso.com/documents/summermarketperformancereportforaugust2023.pdf}{[Online]}.

\bibitem{CAISO2022Assessment}
California ISO,
``2022 Summer Loads and Resources Assessment,''
May 18, 2022, p. 5, Table 1.
\href{https://www.caiso.com/documents/2022-summer-loads-and-resources-assessment.pdf}{[Online]}.

\bibitem{EIA2020Wildfire}
U.S. Energy Information Administration,
``Smoke from California wildfires decreases solar generation in CAISO,''
\emph{Today in Energy}, Sept. 30, 2020.
\href{https://www.eia.gov/todayinenergy/detail.php?id=45336}{[Online]}.

\bibitem{CAISO2024DMM}
California ISO, Department of Market Monitoring,
``2024 Annual Report on Market Issues and Performance,''
Aug. 7, 2025, sec. 12.2, p. 266.
\href{https://www.caiso.com/documents/2024-annual-report-on-market-issues-and-performance-aug-07-2025.pdf}{[Online]}.

\bibitem{Terren2026Forecast}
Guillermo Terrén-Serrano, Ranjit Deshmukh, Manel Martínez-Ramón,
``Probabilistic day-ahead forecasting of system-level renewable energy and electricity demand,''
\emph{Nature Communications}, vol. 17, Art. no. 3307, 2026.
\href{https://doi.org/10.1038/s41467-026-69015-w}{doi: 10.1038/s41467-026-69015-w}.

\end{thebibliography}
\end{document}